\RequirePackage{fix-cm}

\documentclass[aps,twocolumn,superscriptaddress,amsmath,amssymb, noshowpacs]{revtex4}

\usepackage{graphicx}
\usepackage{hyperref}
\usepackage{amsmath}

\newcommand{\ket}[1]{\vert#1\rangle}

\begin{document}

\title{Implementation of a hybrid scheme for coherent plug-and-play quantum key distribution}

\author{Ignacio H. L\'opez Grande}
\affiliation{DEILAP, CITEDEF-CONICET, J.B. de La Salle 4397, 1603 Villa Martelli, Buenos Aires,
Argentina}
\affiliation{Departamento de Física, FCEyN, UBA, Ciudad Universitaria, 1428 Buenos Aires, Argentina}
\author{Miguel A. Larotonda}
\affiliation{DEILAP, CITEDEF-CONICET, J.B. de La Salle 4397, 1603 Villa Martelli, Buenos Aires,
Argentina}
\affiliation{Departamento de Física, FCEyN, UBA, Ciudad Universitaria, 1428 Buenos Aires, Argentina}
\affiliation{UNIDEF, CONICET, Buenos Aires, Argentin}

\date{08 June 2018}

\begin{abstract}
We experimentally demonstrate a hybrid configuration  for Quantum Key Distribution, that combines the simplicity of Distributed Phase Reference protocols with the self-referencing features and polarization insensitivity of the so-called Plug \& Play system.  Additionally,  all the components are arranged in a server-client scheme to allow for practical key distribution. Blank, coherent pulse pair trains are generated at the reception end of the link by means of a pulse sequence and an unbalanced interferometer, and sent to the other end. The emitter writes the qubits by erasing one of the pulses from the pair as in a Coherent-One Way protocol. Detection, as well as  eavesdropping monitoring is performed at the receiver side, using the same interferometer that was used to generate the initial phase-referenced pulses.

\end{abstract}
\maketitle

\section{Introduction}
\label{intro}
Quantum key distribution (QKD) is the name of a collection of techniques and protocols that guarantees secure communication between two parties using the principles of quantum mechanics. Users at both ends of the QKD link produce a shared random secret key, which they can later use to encrypt and decrypt information. The secret key is to be known only to the two communicating users. The laws of quantum mechanics give them the ability to detect the presence of any third party trying to gain knowledge of the key: anyone trying to eavesdrop on the key eventually has to measure it, hence disturbing the system and introducing measurable changes. QKD was originally introduced in 1984 \cite{bennett1984quantum}, and since then it evolved rapidly, particularly in the last two decades. This evolution led to a fast growth of the field, which is now mature enough to start to be implemented in demanding real life environments such as metropolitan fiber networks \cite{peev2009secoqc,stucki2011long,wang2014field,tang2016measurement} and ground-to-satellite quantum links \cite{liao2017satellite}.  

Massive and global implementation of QKD systems demands fast and simple schemes for secret key distribution. Efforts and achievements have been concentrated in improving components \cite{semenov2001quantum,gol2001picosecond} and developing better protocols \cite{muller1997plug,inoue2002differential,stucki2005fast} towards fast, reliable and continuous operation in real world telecommunication networks \cite{wang2014field,maeda2009technologies,poppe2009results,sasaki2017quantum}. The present work attempts to introduce an improvement on a QKD protocol designed to operate on telecom optical fiber, by combining a Plug \& Play interferometric setup that provides self phase referencing and compensation of birefringence  \cite{ribordy1998automated,muller1997plug}, and the simplicity and potential high-speed performance of distributed-phase-reference protocols such as Differential Phase Shift (DPS) \cite{inoue2002differential,takesue2005differential,diamanti2006100,takesue2007quantum} and Coherent One-Way (COW) QKD \cite{gisin2004towards,stucki2005fast,stucki2009high,stucki2009continuous}. Furthermore, the presented setup is arranged in a client-server QKD scheme, where the complex and sensitive resources such as the light source and the detectors are situated at the server side. On the other side, the client setup only comprises an amplitude modulator arranged in a passive Sagnac loop.

The key is generated according to the COW protocol, in which a train of pulses share a constant relative phase, and the bits are encoded in pairs of pulses by varying the pulse intensity. Shared, stable and self aligning phase is obtained by means of a pulse pattern carved within the laser photon's wave packet and an unbalanced interferometer that is used both to generate and to monitor the inter-pulse coherence. This last effect is achieved by adopting the double pass of the Plug \& Play configuration: a train of pulses is produced by Bob and sent to Alice. She writes the bits and sends the pulse train back to Bob, who measures on a data line to obtain the key, and on an interferometric monitoring line to detect an eventual eavesdropper that alters the visibility of the interferometer by breaking the coherence between pulses \cite{gisin2004towards,moroder2012security}.

Alice's stage is implemented with a Sagnac fiber loop based on a polarizing beamsplitter that contains an intensity modulator acting on the two polarization states. The device ensures an optimized efficiency of the modulator, regardless of the state of polarization at the input.

\section{Description of the setup}
\label{sec:setup}

\begin{figure}
\includegraphics[width=0.45\textwidth]{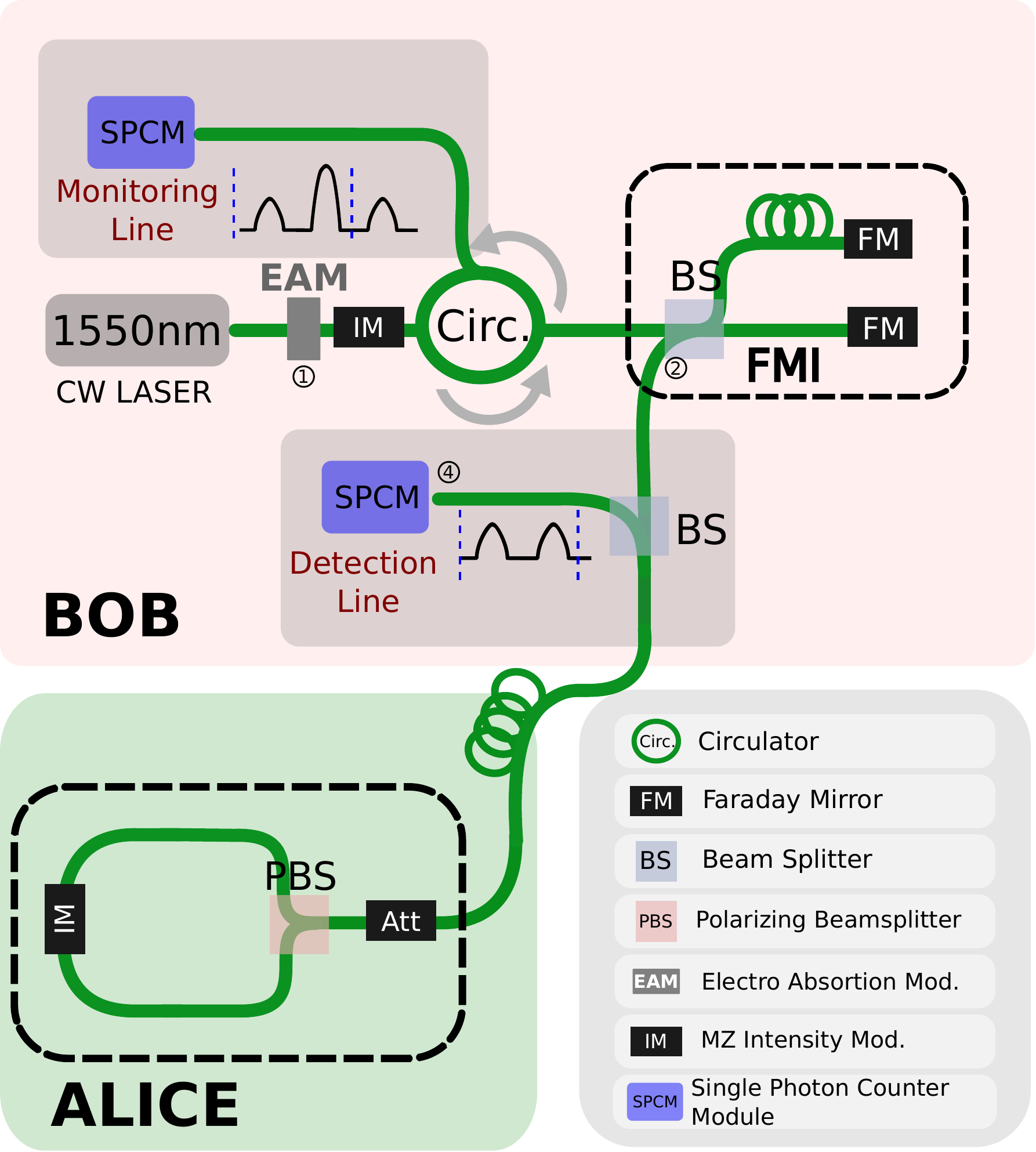}
\caption{\footnotesize{\emph{The hybrid scheme for Coherent Plug \& Play Quantum Key Distribution. The server-side is composed by a CW DFB laser followed by an electro-absorption modulator which generates 800~ps light pulses, a circulator and the FMI which generate intense "blank pulses" patterns. For the detection the server side uses the same FMI to measure the coherence of the received sequence and a tap to detect the arrival time of the photons, both with a single SPCM in a multiplexed scheme (not shown). The client-side consists just in a intensity modulator displayed in a ring configuration by means of a Polarizing Beam Splitter}}}
\label{fig:setup}
\end{figure}

The scheme is based on a Plug \& Play architecture, with distributed phase reference (COW) encoding. In the proposed setup, blank symbols are defined by pairs of pulses, which must have a fixed phase reference. These symbols are generated at Bob's stage using a combination of a temporal pulse pattern and an unbalanced Michelson interferometer with Faraday mirrors, and subsequently sent to Alice. In turn, Alice generates the key by writing the bits on each of the symbols: `0s' and `1s' are encoded by erasing a single pulse from each symbol (Fig. \ref{fig:setup}). She also leaves some symbols untouched (\emph{decoy} states) to monitor for eavesdropping, before sending the stream back to Bob: the possible states are therefore the early state $\ket{e}$, which holds for the key bit value `0', the late state $\ket{\ell}$ corresponding to the bit `1' of the secret key and a coherent superposition of this two, the decoy state:  $\ket{e}+e^{i\phi\left(t\right)}\ket{\ell}$. The relative phase 
$\phi$ is in principle a slowly varying function of time associated to the mechanical and thermal fluctuations of the interferometer. 

In the last part of the process, Bob reads the bits at the main line and measures visibility with his interferometer to check for coherence between successive optical pulses at the monitoring line. An alteration on the interferometric visibility is associated with an eavesdropper attack. Most of the critical and resource demanding tasks are performed at the server side, while the client side only comprises an intensity modulation operation.

\subsection{Pulse pattern generation and qubit writing}
\label{subsec:qubitgen}

\begin{figure}
\includegraphics[width=0.45\textwidth]{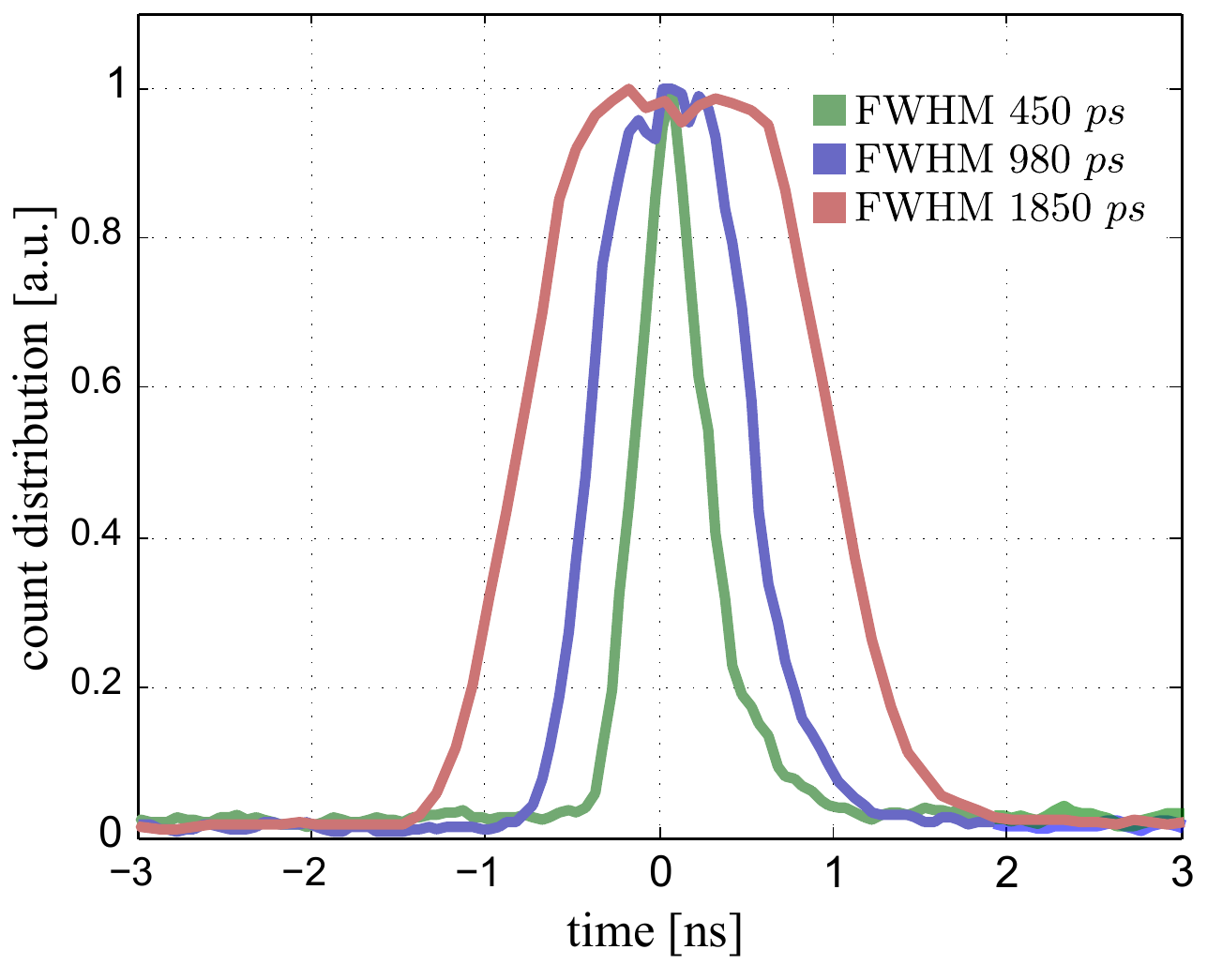}
\caption{\footnotesize{\emph{Temporal count distribution for different light pulse widths obtained by controlling the Electro-Absorption Modulator of the telecom CW laser. The three temporal profiles are generated with different settings of the EAM driver.}}} 
\label{fig:singlepulse}
\end{figure}

Blank symbols are generated as intense light pulses, arranged in 435~MHz pulse packets. In this demonstration, in order to simplify the analysis we use the shortest possible pulse packet, 
consisting of four pulses that share a common phase.
First, a sequence of two pulses separated by 4.6~ns is tailored from a single wavepacket of a 1548~nm DFB laser with an electro-absorption modulator (EAM). 
An \emph{ad-hoc} built electrical pulse generator based on fast ECL logical gates and a commercial EAM driver (Maxim Integrated MAX3941) were used for that purpose;  using this combination 
we are able to generate optical pulses as short as 450~ps FWHM at frequencies up to hundreds of MHz. Figure \ref{fig:singlepulse} shows pulses with different widths obtained by these means. In order to increase the signal-to-noise ratio of the pulses, an Intensity Modulator (IM) generating  35~ns transmission windows was added to the setup.

Finally the pulses enter into a Faraday Michelson Interferometer (FMI), with a length unbalance between the arms that halves the initial pulse separation, \linebreak $\Delta \tau$~=~2.3~ns.  
Four equally spaced pulses are thus obtained at the output of the interferometer. The temporal width of the pulses was electronically set to 800~ps FWHM, which allows for a coarser temporal match between the initial two-pulse sequence separation and the time-of-flight difference in the interferometer.

Despite the polarization not being controlled and the pulses going over different paths, the use of a FMI ensures that the polarization is the same for the four pulses at the output of the interferometer.
This is crucial to get high visibility at the interferometric monitoring stage.
Throughout this process, patterns of four pulses are created at a rate of 115~kHz.
An FPGA engine controls the trigger and separation between pulses, as well as the timing and synchronization for the rest of the protocol hardware.

The quantum state writing takes place at the client-side (Alice) and is accomplished using a simple and robust setup compatible with the plug and play requirements; i.e. it must be polarization insensitive. Contrastingly, commercial Mach-Zehnder intensity modulators (IM) usually have a strong dependence with the polarization.
To overcome this polarization dependence we use a combination of a polarization beamsplitter (PBS) plus an IM, as described in the appendix  \ref{app:ring}. The effect of such arrangement is a polarization-insensitive action of the intensity modulator, plus a polarization rotation. 

\begin{figure}
\includegraphics[width=0.45\textwidth]{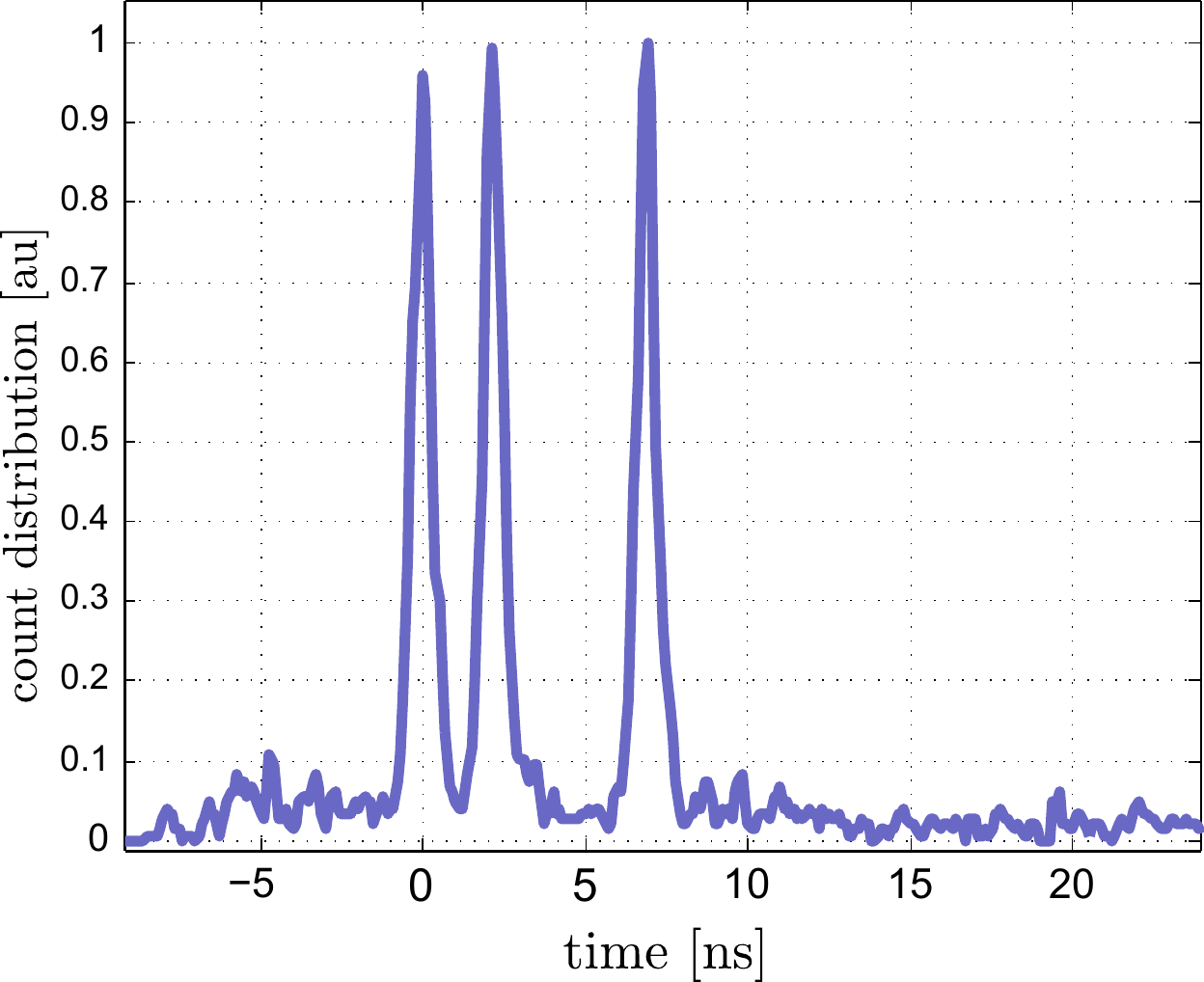}
\caption{\footnotesize{\emph{State $\ket{decoy,\ell}$ prepared by Alice carved from a blank pattern. The state  is prepared using an IM which is synchronized with the arrival time of the blank pulses at the client-side.}}}
\label{fig:state1101}
\end{figure}

For a packet of two symbols, where each symbol can have three values, a total of nine different sequences can be written by Alice.
The IM used to tailor these states is driven by a combination of electrical pulse generator and  similar to the one used for the EAM. In this case we set the width of the electrical pulses to be 1850~ps, in this way we assure a good  extinction of the desired light pulses even with a coarse time synchronization provided by the FPGA (steps of $\sim$ 800 ps).  
Figure \ref{fig:state1101} shows one of the prepared sequences ( $\ket{decoy,\ell}$ ) at Alice's side. 

\subsection{Qubit reading and monitoring lines}
\label{subsec:qubitread}

\begin{figure}
\includegraphics[width=0.45\textwidth]{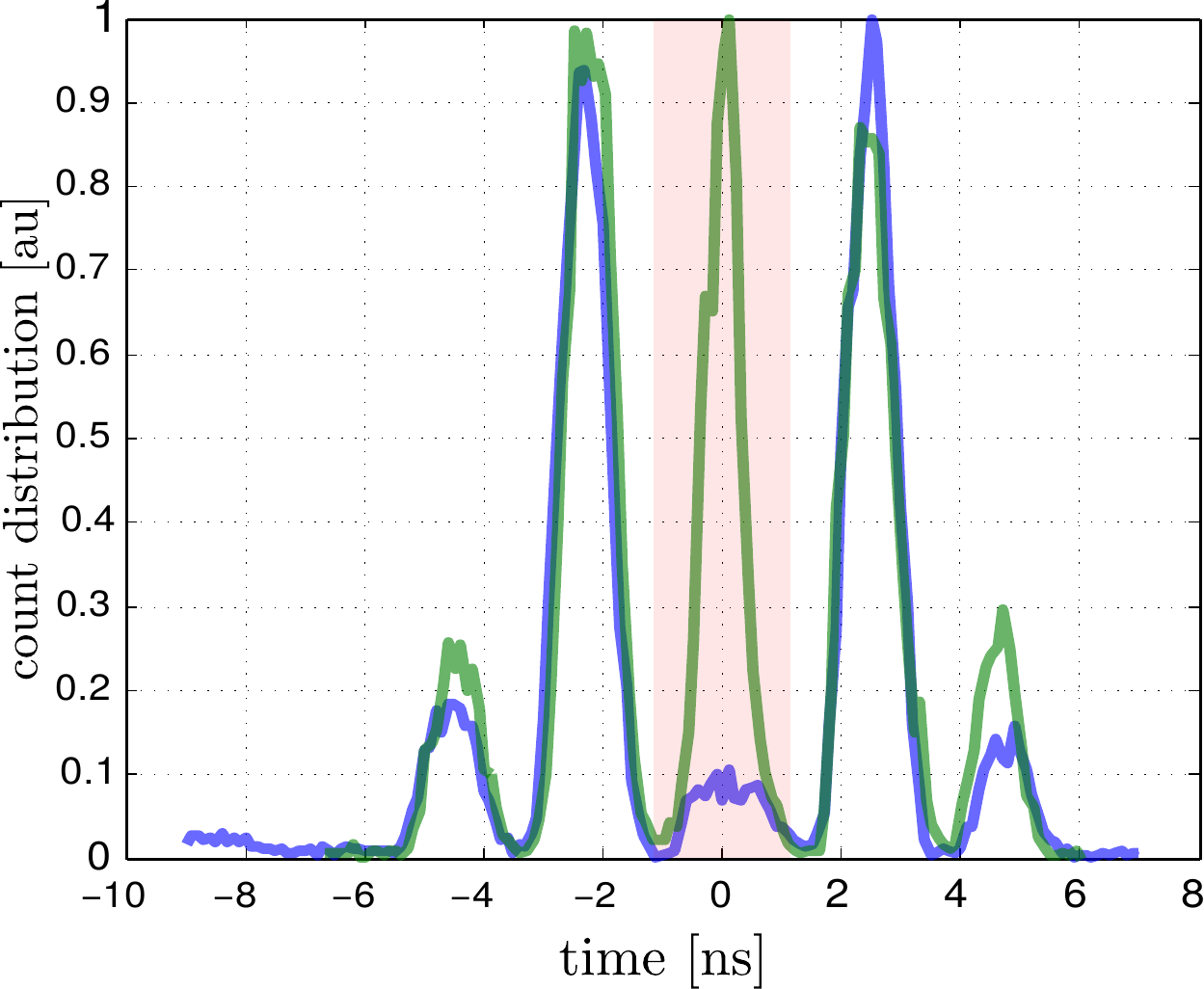}
\caption{\footnotesize{\emph{Temporal distribution showing the  $\ket{decoy,decoy}$ state detected with the monitoring line at two different situations. The green (lighter) curve shows constructive interference at the central peak, while the violet (darker) curve shows a destructive interference condition.}}}.
\label{fig:interf_max_min}
\end{figure}

The detection of the quantum states is performed at the server side in two complementary bases, the  monitoring line and the data line. The latter registers the arrival time of the detected photons which encode the bits of the secret key, but gives no information when detecting decoy states. 
In addition, the monitoring line measures coherence between subsequent pulses. For this purpose, the pulses re-enter the unbalanced FMI. Pulses originally separated $\Delta\tau$ will interfere at the FMI output and guided to the monitoring detector via an optical circulator. 
Interference can be evaluated differently depending on the arrival time of the detected photons. Coherence between pulses is broken when sophisticated coherent attacks are performed over several pulses, and it can be measured between pulses belonging  either to the same symbol or to adjacent symbols: interference between pulses that pass once through each of the arms of the FMI is called ``intra-bit'' interference; in this case paths are automatically balanced and light exits through the desired output of the beam splitter if an untampered decoy pulse has been measured. In contrast, ``across-the bit'' interference, 
which arises from two adjacent pulses of consecutive symbols, the early pulse passing two times through the long path and the late pulse going twice through the short one) will be recombined with an arbitrary phase depending on the path difference, hence requiring some kind of stabilization procedure to assure the correct 
measure of coherence between subsequent laser pulses. Figure \ref{fig:interf_max_min} shows the count distribution for the state $\ket{decoy,decoy}$ with no active stabilization when constructive and destructive interference takes place in the monitoring line. With these two conditions we can estimate the visibility of the interferometer. Nevertheless, when using longer pulse patterns, Alice and Bob can agree to shift the pulse sequence by a single pulse, thus turning the ``intra-bit'' pulse separation into an ``across-the bit'' pulse separation.

In this proof-of-concept experiment, we use only one SPCM to perform the detection in both bases by temporally multiplexing the light signals with a beamsplitter. By tuning the splitting ratio of the BS at the detection line the probability of data and monitoring line detection selection bases can be chosen: we have intentionally set the monitoring line to data line detection probability to be balanced to show more clearly the results of the two bases within the same data stream. In a real world implementation, though, the portion of the signal that is sent to the monitoring line is in the order of 10\%.

\begin{figure}
\includegraphics[width=0.45\textwidth]{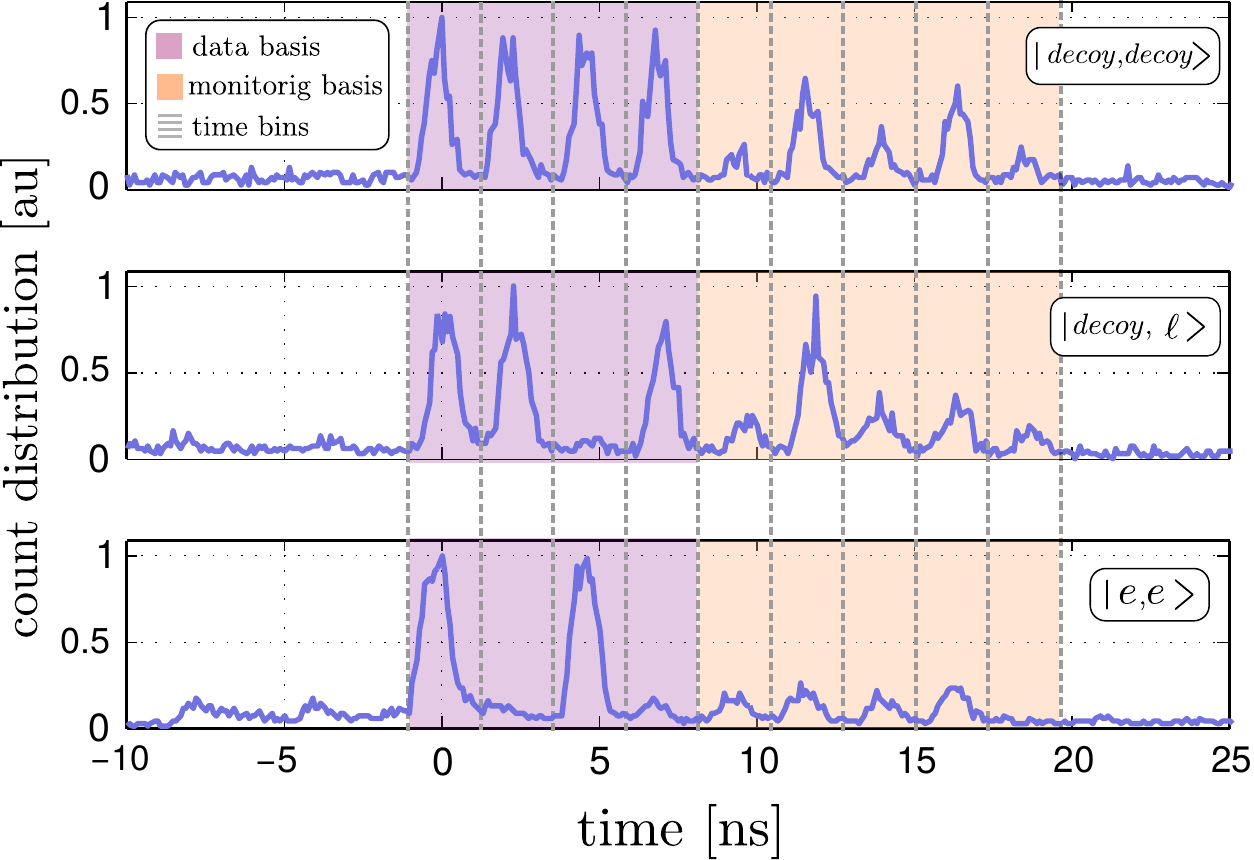}    
\caption{\footnotesize{Count distribution for three different states detected at the server-side. The two detection bases are time multiplexed to use only one SPDM. The orange and purple areas correspond to the monitoring line and data line time-bins respectively. The detection basis has a higher detection rate with a $65\%$ of the detected photons.}}
\label{fig:patt}
\end{figure}

Figure \ref{fig:patt} shows the count distribution at server side for three states prepared by Alice, with different number of decoy states. Single photon detection on either the detection and the monitoring line takes place by dividing the signal into 2.3 ns time bins. 

We calculate an average QBER of $6\%$ from the different states that can be prepared. The origin of this noise comes from two main sources: on the one hand the extinction ratio of the IMs (Agere 2623N) is around 13~dB for RF driving signals. This limited performance at the qubit writing stage accounts for almost $80\%$ of our mean QBER value. On the other hand, there is an increased background noise in the multiplexed detection stage we use for this demonstration: when preparing the blank symbols part of the intense light at the output of the FMI is coupled into the SPCM and therefore appears as background noise in the monitoring and data detection basis. We tackle this issue by inserting an extra IM after the laser output, which generates 35~ns  transmission windows synchronized with the output pulses. This mitigates the amount of spurious light at the SPCM, although the achieved extinction ratio is not high enough to neglect this effect. The system was run in a condition where the measured mean photon rate per symbol at the detection stage is $R=0.01$; in this condition, the detector noise and the background light contribute with the remaining portion or the bit errors. 

The visibility of the decoy states is obtained from the measurements in the monitoring line. For the self-stabilized peaks (figure \ref{fig:interf_max_min}) we measure a visibility of  $0.95\pm0.05$. We also study the fluctuations of the unstabilized interference peak. Figure \ref{fig:interf_max_min} shows the count rate for the central peak when Alice is emitting a \linebreak $\ket{decoy,decoy}$ state and no active compensation is performed attaining a visibility value of $0.93\pm0.03$. The reduced visibility is mainly due to the different arrival time (around 100~ps) between the two pulses at the output of the FMI.

\section{Discussion}
\label{sec:discuss}

Regarding the security of the protocol, it was shown that Photon Number Splitting (PNS) attacks are not effective in the COW protocol, since there is information encoded in the phase difference between pulses, so any PNS attack will break
the sequential coherent pulses, resulting in an alteration of the visibility in Bob's monitoring line measurements.
General security bounds against individual attacks and also upper bounds for the error rates against coherent attacks have been derived for the COW protocol \cite{branciard2008upper}; in the same work upper bounds for collective beam splitting attacks were also obtained. Zero-error attacks have also been studied for the COW protocol \cite{branciard2007zero}. More recently, lower bounds on the key generation rate in a finite-size key scenario have been obtained \cite{moroder2012security}. 

In the present setup, additional considerations must be taken into account due to the plug and play configuration: during the transmission from Bob to Alice, coherent, multiphoton pulses may be easily tapped by Eve, but there is no encoded information at this stage. 
Once the bits are written by Alice and the pulse stream is sent back to Bob, intense pulses that reach the client side must be dimmed down to the single photon level before being sent through the quantum channel. The insertion loss of Alice's arrangement is only 9~dB. Further attenuation is achieved using an optical attenuator (Fig. 1), which imposes a total loss of 46 dB in its double pass setting. Finally another 2 dB of attenuation is produced by the two meter FC/PC fiber connectorized patchcord separating Alice and Bob. It is worth to note that the detection line is placed immediately at the server input, and therefore the only additional losses are imposed by the optical tap that sends a small portion of the light to the monitoring line (with a transmittance as high as 0.9).

In this way, this particular demonstration starts with a blank pattern of intense light pulses, which is lowered down to a desired mean photon number output per pulse at the client side of $\mu_A$=0.5 \cite{stucki2009high}. We obtain a mean photon number at the detector of $\mu_B$=0.1 (a measured photon rate $R$=0.01 taking into account a 10\% efficiency of the single photon counter). Such attenuation can be interpreted to be equivalent to a 7 dB link loss, that is, roughly 30 km of single mode fiber. Eavesdropping is now meaningful, but as discussed above, it is obtained at the expense of breaking the coherence between pulses and reducing the visibility at the monitoring line. The protocol is therefore sensitive against intercept-resend attacks and photon-number-counting attacks performed coherently on two subsequent pulses \cite{gisin2004towards}.

We can estimate an upper bound for the secret key rate, based on a security analysis that considers collective beam splitting attacks \cite{branciard2008upper}. The security analysis is similar to the one performed in a standard COW protocol \cite{stucki2009high}, and in other, more recent, distributed-phase-reference protocols \cite{bacco2016two}. Assuming an interferometer visibility of $V$=0.93, a dark count probability of $2.5\times10^{-5}$ per time-bin, a ratio of key-to-decoy state emission of 9:1 and an input mean photon rate per pulse of $\mu_A$=0.5, for a 30 km fiber link we obtain a sifted key rate of $R_{sift}=8.9\times10^{-3}$ which is in very good agreement with the measured rate, and a theoretical secure key rate of $R_{sec}=1.8\times10^{-3}$ symbols per pulse. Optimization of the input photon rate for such link attenuation leads to a maximum $R_{sec}=2.0\times10^{-3}$, for an optimized input photon rate per pulse of $\mu_A$=0.38. 
In the described experimental conditions, doubling the link length (from 30 km to 60 km) implies a reduction of the secure key rate by an order of magnitude. For such large distances, Eve's available information is the term that dominates the reduction of the secure key extraction rate.

Finally, it is worth mentioning that it has been also demonstrated that non-idealities that are present in any physical implementations of QKD can be practically exploitable to obtain information on the generated key \cite{gerhardt2010full}. The amount of light at the output of the client must be monitored to prevent a side channel attack (i.e. Trojan Horse attack \cite{gisin2006}).  This task can be accomplished using a PIN photodiode placed at the input of the client setup and a narrow band pass filter. We have explicitly omitted this detector to ease the description of the quantum channel.
 Additionally, a linear optical detector might be included at the server side to monitor the presence of intense light pulses designed by an eavesdropper to perform a faked-state-type attacks on the counting device \cite{makarov2005faked}.

\section{Conclusions}
\label{sec:conclusion}
We have presented a variation of the Coherent One-Way QKD protocol in a Plug \& Play configuration. This arrangement allows for the utilization of a self-referenced interferometer for the monitoring line detection, and for the deployment of all the sensitive and costly hardware components at a server side, while maintaining only an intensity modulator at the client side. The pulse pattern, synchronization, qubit writing and detector triggering were achieved using electronic pulsers built from ECL logic components and an overclocked FPGA board as the timing engine. The visibility of the unperturbed intra-bit interference is stable and above 95\%, and a 6\% QBER. The main feature of the present scheme is the efficient use of the resources and the robustness of the interferometric monitoring line, making it suitable for operation under realistic environmental conditions.

\begin{acknowledgements}
We acknowledge financial support from CONICET and ANPCyT funding agencies, and the \mbox{PIDDEF} program from MINDEF. 
We thank L. Knoll for fruitful discussions. 
\end{acknowledgements}

\appendix

\section*{Appendix: PBS Ring Intensity Modulator}
\label{app:ring}

As mentioned in section \ref{sec:setup} we implement a Polarization Insensitive Intensity Modulator (PIIM) by using an IM and a PBS in a ring configuration.

The PBS separates the incoming light into two orthogonal (linear) polarizations, which are coupled into the same propagation mode (slow axis of the PM fibers) at the PBS output ports. The IM is inserted symmetrically between the two PBS ports, forming a closed, bidirectional loop that connects the two ports of the PBS. 

\begin{figure}
\includegraphics[width=0.45\textwidth]{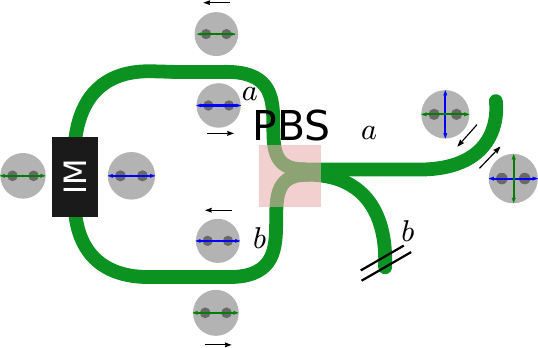}
\caption{\footnotesize{\emph{A Polarization Insensitive Intensity Modulator is achieved by exploiting the PBS mode coupling in polarization maintaining fibers. This schematic diagram shows how the two modes enter to the PIIM at the same time with different propagation directions thus avoiding any polarization effect on the writing process.}}}
\label{fig:ring}
\end{figure}

The PBS has input and output polarization-maintaining Panda type optical fibers. Defining the two fiber propagation modes of orthogonal polarization as $\ket{H}$ and $\ket{V}$ (linear polarization states), the state of a photon entering the PBS can be generally described as: 
   $\ket{\psi_{in}}=\alpha \ket{H} + \beta \ket{V}$
where $\alpha$, $\beta$ fulfill the standard normalization criteria. The PBS couples each incoming polarization component into the different ports with the same output polarization; these outputs are then connected to both ports of the IM.
By doing this we ensure that all the light received by Alice circulates through the IM with the adequate polarization state, allowing for maximum modulation efficiency: the states are written on the pulses by the IM acting simultaneously on both signals. 
At the end of the trip through the ring, the PBS ports are swapped (output $b$($a$) with input $a$ ($b$), figure \ref{fig:ring}). 
The overall  polarization transformation produced by the PIIM can be stated in the following way:

\begin{equation}
PIIM
\left\lbrace
\begin{array}{ll}
\ket{H}\rightarrow{}\,\,\,\,\,\,\ket{V}\\
\ket{V}\rightarrow{}-\ket{H},\\
\end{array}
\right.
\end{equation}

obtaining an output state $\ket{\psi_{out}}=\alpha \ket{V}- \beta \ket{H}$.
This transformation resembles the one performed by a Faraday Mirror, although all the rotations are reciprocal: PIMM $\propto\sigma_y$, where $\sigma_y$ is the Pauli matrix which has $\ket{R}$ and $\ket{L}$ as eigenvectors.  The invariant states of such ``reflection'' are the circular polarization states, whereas a Faraday mirror action is proportional to $\sigma_x$  \cite{QFMirror}, with the diagonal states as eigenvectors.
As a side result, the counter-propagating pulse trains are recombined at the PBS into a state which has its polarization rotated 90$^{\circ}$ with respect to the input state. 

It is worth to notice that the device can be converted into a ring Faraday mirror by adding an in-line $90\deg$ Faraday rotator with its ports coupled to different axis of polarization maintaining pigtails. In this situation, any alteration to the polarization state induced by thermal and mechanical perturbations of the fiber link are minimized.
  


\begin{thebibliography}{10}
\providecommand{\url}[1]{{#1}}
\providecommand{\urlprefix}{URL }
\expandafter\ifx\csname urlstyle\endcsname\relax
  \providecommand{\doi}[1]{DOI \discretionary{}{}{}#1}\else
  \providecommand{\doi}{DOI \discretionary{}{}{}\begingroup
  \urlstyle{rm}\Url}\fi

\bibitem{bennett1984quantum}
C.H. Bennett, G.~Brassard, in \emph{International Conference on Computers,
  Systems \& Signal Processing, Bangalore, India, Dec 9-12, 1984} (1984), pp.
  175--179

\bibitem{peev2009secoqc}
M.~Peev, C.~Pacher, R.~All{\'e}aume, C.~Barreiro, J.~Bouda, W.~Boxleitner,
  T.~Debuisschert, E.~Diamanti, M.~Dianati, J.~Dynes, et~al., New Journal of
  Physics \textbf{11}(7), 075001 (2009)

\bibitem{stucki2011long}
D.~Stucki, M.~Legre, F.~Buntschu, B.~Clausen, N.~Felber, N.~Gisin, L.~Henzen,
  P.~Junod, G.~Litzistorf, P.~Monbaron, et~al., New Journal of Physics
  \textbf{13}(12), 123001 (2011)

\bibitem{wang2014field}
S.~Wang, W.~Chen, Z.Q. Yin, H.W. Li, D.Y. He, Y.H. Li, Z.~Zhou, X.T. Song, F.Y.
  Li, D.~Wang, et~al., Optics Express \textbf{22}(18), 21739 (2014)

\bibitem{tang2016measurement}
Y.L. Tang, H.L. Yin, Q.~Zhao, H.~Liu, X.X. Sun, M.Q. Huang, W.J. Zhang, S.J.
  Chen, L.~Zhang, L.X. You, et~al., Physical Review X \textbf{6}(1), 011024
  (2016)

\bibitem{liao2017satellite}
S.~Liao, W.~Cai, W.~Liu, L.~Zhang, Y.~Li, J.~Ren, J.~Yin, Q.~Shen, Y.~Cao,
  Z.~Li, et~al., Nature \textbf{549}(7670), 43 (2017)

\bibitem{semenov2001quantum}
A.D. Semenov, G.N. Gol'tsman, A.A. Korneev, Physica C: Superconductivity
  \textbf{351}(4), 349 (2001)

\bibitem{gol2001picosecond}
G.~Gol'tsman, O.~Okunev, G.~Chulkova, A.~Lipatov, A.~Semenov, K.~Smirnov,
  B.~Voronov, A.~Dzardanov, C.~Williams, R.~Sobolewski, Applied Physics Letters
  \textbf{79}(6), 705 (2001)

\bibitem{muller1997plug}
A.~Muller, T.~Herzog, B.~Huttner, W.~Tittel, H.~Zbinden, N.~Gisin, Applied
  Physics Letters \textbf{70}(7), 793 (1997)

\bibitem{inoue2002differential}
K.~Inoue, E.~Waks, Y.~Yamamoto, Physical Review Letters \textbf{89}(3), 037902
  (2002)

\bibitem{stucki2005fast}
D.~Stucki, N.~Brunner, N.~Gisin, V.~Scarani, H.~Zbinden, Applied Physics
  Letters \textbf{87}(19), 194108 (2005)

\bibitem{maeda2009technologies}
W.~Maeda, A.~Tanaka, S.~Takahashi, A.~Tajima, A.~Tomita, IEEE Journal of
  Selected Topics in Quantum Electronics \textbf{15}(6), 1591 (2009)

\bibitem{poppe2009results}
A.~Poppe, T.~L{\"a}nger, T.~Lor{\"u}nser, O.~Maurhart, C.~Pacher, M.~Peev, in
  \emph{European Quantum Electronics Conference} (Optical Society of America,
  2009), p. ED5\_1

\bibitem{sasaki2017quantum}
M.~Sasaki, Quantum Science and Technology \textbf{2}(2), 020501 (2017)

\bibitem{ribordy1998automated}
G.~Ribordy, J.D. Gautier, N.~Gisin, O.~Guinnard, H.~Zbinden, Electronics
  Letters \textbf{34}(22), 2116 (1998)

\bibitem{takesue2005differential}
H.~Takesue, E.~Diamanti, T.~Honjo, C.~Langrock, M.~Fejer, K.~Inoue,
  Y.~Yamamoto, New Journal of Physics \textbf{7}(1), 232 (2005)

\bibitem{diamanti2006100}
E.~Diamanti, H.~Takesue, C.~Langrock, M.~Fejer, Y.~Yamamoto, Optics Express
  \textbf{14}(26), 13073 (2006)

\bibitem{takesue2007quantum}
H.~Takesue, S.W. Nam, Q.~Zhang, R.H. Hadfield, T.~Honjo, K.~Tamaki,
  Y.~Yamamoto, Nature photonics \textbf{1}(6), 343 (2007)

\bibitem{gisin2004towards}
N.~Gisin, G.~Ribordy, H.~Zbinden, D.~Stucki, N.~Brunner, V.~Scarani, arXiv
  preprint quant-ph/0411022  (2004)

\bibitem{stucki2009high}
D.~Stucki, N.~Walenta, F.~Vannel, R.T. Thew, N.~Gisin, H.~Zbinden, S.~Gray,
  C.~Towery, S.~Ten, New Journal of Physics \textbf{11}(7), 075003 (2009)

\bibitem{stucki2009continuous}
D.~Stucki, C.~Barreiro, S.~Fasel, J.D. Gautier, O.~Gay, N.~Gisin, R.~Thew,
  Y.~Thoma, P.~Trinkler, F.~Vannel, et~al., Optics Express \textbf{17}(16),
  13326 (2009)

\bibitem{moroder2012security}
T.~Moroder, M.~Curty, C.C.W. Lim, H.~Zbinden, N.~Gisin, et~al., Physical Review
  Letters \textbf{109}(26), 260501 (2012)


\bibitem{branciard2008upper}
C.~Branciard, N.~Gisin, V.~Scarani, New Journal of Physics \textbf{10}(1),
  013031 (2008)

\bibitem{branciard2007zero}
C.~Branciard, N.~Gisin, N.~Lutkenhaus, V.~Scarani, Quantum Information \&
  Computation \textbf{7}(7), 639 (2007)

\bibitem{bacco2016two}
D.~Bacco, J.~Bjerge, M.~Usuga Castaneda, Y.~Ding, S.~Forchhammer, K.~Rottwitt, L.~Oxenl{\o}we, Scientific Reports \textbf{6}, 36756 (2016)

\bibitem{gerhardt2010full}
I.~Gerhardt, Q.~Liu, A.~Lamas-Linares, J.~Skaar, C.~Kurtsiefer, V.~Makarov,
  arXiv preprint arXiv:1011.0105  (2010)

\bibitem{gisin2006}
N.~Gisin, S.~Fasel, B.~Kraus, H.~Zbinden, and G.~Ribordy
Physical Review A \textbf{73}, 022320 (2006)

\bibitem{makarov2005faked}
V.~Makarov, D.R. Hjelme, Journal of Modern Optics \textbf{52}(5), 691 (2005)

\bibitem{QFMirror}
V.~Secondi, F.~Sciarrino, F.~De Martini, Physical Review A (70), 040301(R) (2004)

\end{thebibliography}
\end{document}